\documentstyle[aps,prd]{revtex}

\newcommand{\bq}{\begin{equation}}
\newcommand{\eq}{\end{equation}}
\newcommand{\bqn}{\begin{eqnarray}}
\newcommand{\eqn}{\end{eqnarray}}
\newcommand{\nb}{\nonumber}
\newcommand{\lb}{\label}

\begin{document} 
\title{ Gravitational Collapse of 
Perfect Fluid in N-Dimensional Spherically Symmetric Spacetimes}
\author{Jaime  F. Villas da Rocha \thanks{E-mail: roch@on.br}}
\address{ Departmento de Astronomia Gal\'atica e Extra-Gal\'atica, 
 Observat\'orio Nacional~--~CNPq,
 Rua General Jos\'e Cristino 77, S\~ao
Crist\'ov\~ao, 
 20921-400 
Rio de Janeiro~--~RJ, Brazil }
\author{ Anzhong Wang\thanks{E-mail: wang@dft.if.uerj.br}}
\address{  Departamento de F\' {\i}sica Te\' orica,
Universidade do Estado do Rio de Janeiro,
 Rua S\~ ao Francisco Xavier 524, Maracan\~ a,
20550-013 Rio de Janeiro~--~RJ, Brazil}

\date{November 12, 1999}

\maketitle

\begin{abstract}

The Riemann, Ricci and Einstein tensors for N-dimensional spherically
symmetric spacetimes in various systems of coordinates are studied, and  the
general metric for conformally flat spacetimes is given. As an application,
all the Friedmann-Robertson-Walker-like solutions for a perfect fluid with an
equation of state  $p = k \rho$  are found.  Then, these solutions are used to
model the gravitational collapse of a compact ball, by first cutting them 
along a timelike   hypersurface   and then joining them with  asymptotically
flat Vaidya solutions. It is found that when the collapse has continuous
self-similarity, the formation of black holes always starts with zero mass,
and when the collapse has no such  symmetry, the formation of black holes
always starts with a finite non-zero mass. 
\end{abstract}

\vspace{.6cm}


\section{ Introduction}

Critical phenomena in gravitational collapse have attracted much 
attention \cite{Gu1997} since the pioneering work of Choptuik
\cite{Ch1993}. From the known results obtained so far, the following
emerges \cite{WO1997}: Critical collapse of a
fixed matter field in general can be divided into three different classes
according to the self-similarities that the critical solution possesses. 
If the critical solution has no self-similarity, continuous or discrete,
the formation of black holes always starts with a  mass gap
(Type I collapse), otherwise it will start with zero mass 
(Type II collapse), and the mass of black holes takes
 the scaling form 
$$
M_{BH} \propto (P - P^{*})^{\gamma},
$$
where $P$ characterizes the strength
of the initial data.
In the latter case, the collapse can be further
divided into two subclasses according to whether 
the critical solution has continuous self-similarity (CSS)  
or discrete self-similarity (DSS). Because of this difference, 
the exponent $\gamma$ is usually
also different. Whether the critical solution is CSS, DSS, or none of
them, depending on both the matter field
and the regions of the initial data space 
\cite{Gu1997}.  The co-existence of Type I and Type II collapse 
was first found in the SU(2) Einstein-Yang-Mills case \cite{CCB1996},
and later extended to both the Einstein-scalar case \cite{CLH1997}
and the Einstein-Skyme case \cite{BC1998},
while the co-existence of CSS and DSS critical solutions was 
found in the Brans-Dicke theory \cite{LC1996}. 

The uniqueness of the
exponent $\gamma$ in Type II collapse is well understood in terms of
perturbations  \cite{HKA1996}, and is closely related to the fact that the
critical  solution has only one unstable mode. This property now is considered
as  the main criterion for a solution to be critical \cite{Gu1997}.
While the uniqueness of the exponent $\gamma$ crucially 
depends on the numbers of the unstable
modes of the critical solution, that {\em whether or not the formation
of black holes starts with a mass gap  
seemingly only depends on whether
the spacetime has self-similarity or not}. Thus, even the
collapse is not critical, if a spacetime has CSS or DSS,
 the formation of black holes may still turn on
with zero mass.   To study this problem
in its general term, it is found difficult. Recently, we studied it
for gravitational
collapse of massless scalar field and radiation fluid \cite{WRS1997} and
lately extended it to the case of perfect fluid \cite{RWS1999},  and
 found that when solutions have CSS, the formation of black holes indeed
starts with zero-mass, while when solutions have no  such symmetry it starts
with a mass gap. 

Lately, there has been interest in studying critical collapse in higher
dimensional spacetimes \cite{SH1996}. In particular, it was found that the
exponent $\gamma$ depends on the dimensions of the spacetimes considered
\cite{GCD1999}. This is similar to the critical phenomena in Quantum Field 
Theory and Statistic Mechanics \cite{Goldenfeld1992}. 

In this paper, we shall generalize our previous studies to the case of perfect
fluid in N-dimensional spacetimes. Specifically, in Sec.II we shall 
derive the general metric for comformally flat spherically symmetric
spacetimes. As an application, all the Friedmann-Robertson-Walker-like (FRW)
solutions for a perfect fluid  with a state equation $p = k \rho$ are found,
where $\rho$ is the energy density of the fluid, $p$ the pressure, and $k$ an
arbitrary constant. In Sec. III the main properties of these solutions are
studied in the context of gravitational collapse, and it is found that some of
these solutions represent the formation of black holes, due to the
gravitational collapse of the perfect fluid. However, the mass of such formed
black holes is usually infinitely large. To remend this shortage, in Sec. IV,
the spacetimes are cut along a timelike hypersurface, and then joined to an
asymptotically flat Vaidya solution in N-dimensional spacetimes, so the
resulted black holes have finite masses. Sec. V contains our main conclusions,
while in Appendix A all the physical quantities, such as, the Christoffel
symbols, the Riemann, Ricci, and Einstein tensors, are given in terms of the
two metric, $g_{ab}$, which is orthogonal to the $(N-2)$-dimensional unit
sphere. In Appendix B, the  Christoffel symbols, the Riemann, Ricci and
Einstein tensors in the (1+1)-dimensional spacetimes, $g_{ab}$, and the
extrinsic curvature of a timelike hypersurface, are given in the three
usually used systems of coordinates, the Schwarzschild-like coordinates,
Eddington-Finkelstein-like coordinates, and the double null coordinates.

\section{ The General Conformally flat Metric and the FRW solutions in
N-dimensional Spacetimes}

The general metric for N-dimensional spacetimes with spherical symmetry can
be split into two blocks,
\bq
\lb{eq1}
ds^{2} = g_{ab}(x^{0},x^{1})dx^{a}dx^{b}   -
C^{2}(x^{0},x^{1})d\Omega^{2}_{N-2}, \;(a, b = 0, 1),
\eq
where  $\{x^{\mu}\}\equiv \{x^{0}, x^{1}, \theta^{2}, \theta^{3}, ...,
\theta^{N-1}\}\; (\mu = 0, 1, 2, ..., N-1)$ are  the usual
N-dimensional spherical coordinates,   $d\Omega^{2}_{N-2}$ is the line element
on the unit (N-2)-sphere, given by \bqn
\lb{eq2}
d\Omega^{2}_{N-2} &=& \left(d\theta^{2}\right)^{2} +
\sin^{2}(\theta^{2})\left(d\theta^{3}\right)^{2} + 
\sin^{2}(\theta^{2})\sin^{2}(\theta^{3})\left(d\theta^{4}\right)^{2}\nb\\
& & + ... + \sin^{2}(\theta^{2})\sin^{2}(\theta^{3}) ...
\sin^{2}(\theta^{N-2})\left(d\theta^{N-1}\right)^{2}\nb\\
& = &  \sum^{N-1}_{i = 2}{\left[
\prod^{i-1}_{j =2}\sin^{2}(\theta^{j})\right]
\left(d\theta^{i}\right)^2}.
\eqn
The corresponding physical quantities, such as,
the Christoffel symbols, Riemann, Ricci, and Einstein tensors, are
given in Appendix A in terms of the ($1+1$)-metric, $g_{ab}$. 

It can be shown that  with a
perfect fluid as source, the Einstein field equations, $G_{\mu\nu}
= \kappa \left[(\rho + p)u_{\mu}u_{\nu} - p g_{\mu\nu}\right]$, in
N-dimensional spacetimes can be written as 
\bqn 
\lb{eq3} 
\left(G^{0}_{0} \right.&-& \left.G^{2}_{2}\right)
\left(G^{1}_{1} - G^{2}_{2}\right) -
G^{0}_{1}G^{1}_{0}   = 0, \\ 
\lb{eq4} 
 \rho &=&  \kappa^{-1}\left(G^{0}_{0} + G^{1}_{1} - G^{2}_{2}\right), \\ 
\lb{eq5} 
 p &=&   - \kappa^{-1}  G^{2}_{2}, \\ 
\lb{eq6} 
u_{0}^{2} &=&  \frac{g_{00}(G^{0}_{0} - G^{2}_{2})} 
{G^{0}_{0} + G^{1}_{1} - 2 G^{2}_{2}}, \\
\lb{eq7}
u_{i} &=& 0,  \; (i =  2, 3, 4, ... N-1),
\eqn 
where $\kappa [\equiv 8\pi G/c^{4}]$ is the Einstein constant, $u_{\mu}$ is the
velocity of the fluid. Once $u_{0}$ is known, the component $u_{1}$ can be
obtained from the condition $u_{\lambda}u^{\lambda} = 1$. It is interesting to
note that these equations were first found by Walker in
four-dimensional  spacetime  \cite{Walker1935}. However, the above shows
that  they are  valid even in N-dimensions, and
the dimensional dependence  of the Einstein field equations appear explicitly
only when they are written  in terms of the Ricci tensor,  
\bqn  
\lb{eq8} 
(R^{0}_{0} &-& R^{2}_{2})(R^{1}_{1} - R^{2}_{2}) - R^{0}_{1}R^{1}_{0}  
= 0, \\ 
\lb{eq9} 
\rho &=&  \frac{1}{2\kappa} \left(R^{0}_{0} + R^{1}_{1} - N
R^{2}_{2}\right), \\  
\lb{eq10} 
 p &=&  \frac{1}{2\kappa}\left[R^{0}_{0} + R^{1}_{1} - (N-4) R^2_2 
\right], \\  
\lb{eq11} 
u_{0}^{2} &=&  \frac{g_{00}(R^{0}_{0} - R^{2}_{2})} 
{R^{0}_{0} + R^{1}_{1} - 2 R^{2}_{2}},\\
\lb{eq12}
u_{i} &=& 0,\; \;(i = 2, 3, 4,  ... N-1). 
\eqn 
 
Making the coordinate transformations, $x^{0} = x^{0}({t}, {r}),\;
x^{1} = x^{1}({t}, {r})$, the metric
(\ref{eq1})  can be brought into its isotropic form,
\bq 
\lb{eq13} 
ds^{2} = G(t, r) dt^{2} - K(t, r)\;\left(dr^{2}  
+ r^{2}d\Omega^{2}_{N-2}\right). 
\eq 
Then, it can be shown that the conformally flat condition $C_{\mu\nu\lambda\delta} =0$, 
where $C_{\mu\nu\lambda\delta}$ denotes the Weyl tensor, reduces to a single
equation
\bq
\lb{eq14} 
D,_{rr} - \frac{D_{,r}}{r} = 0, 
\eq
where $D \equiv (G/K)^{1/2}$, and $(\;),_{r} \equiv \partial(\; 
)/\partial r$, etc. The above equation has the general solution
\bq
\lb{eq15}
D(t, r) = f_{1}(t) + f_{2}(t) r^{2},
\eq
where $f_{1}$ and $f_{2}$ are two arbitrary functions of $t$. Hence,   
 there exist three possibilities, 
\bq
\lb{eq16}
i)\; f_{1}(t)\not= 0, f_{2}(t) =
0, \;\;\; ii)\; f_{1}(t)= 0, f_{2}(t) \not= 0,\;\;\;
iii)\; f_{1}(t)\not=
0, f_{2}(t) \not= 0.
\eq
In case i), by introducing a new coordinate $\bar{t} \equiv
\int{f_{1}(t) dt}$ we can bring the metric to a form that is
conformally flat to the Minkowski metric.   Thus, without loss of
generality, in this case we can set $f_{1}(t) = 1$. By a similar
argument, we can set $f_{2}(t) = 1$ in cases ii) and iii). Once this is
done, cases i) and ii) are not independent. In fact,  by a coordinate
transform $r = 1/\bar{r}$, the metric of case ii) will reduce to that of
case i).  Therefore, it is concluded that {\em the  general
conformally flat N-dimensional metric with spherically symmetry}
takes  the form
 \bq
\lb{eq17}
ds^{2} = G(t, r) \left[ dt^{2} - h^{2}(t, r)\;\left(dr^{2} 
+ r^{2}d\Omega^{2}_{N-2}\right)\right],
\eq
where 
\bq
\lb{eq18}
h(t, r) = \left\{
\begin{array}{c}
1, \\
{\left[f_{1}(t) + r^{2}\right]}^{-1},
\end{array}
\right.
\eq
with $f_{1}(t) \not= 0$. In the following, we shall refer solutions
with $h(t,r) = 1$ as Type $A$ solutions, and solutions with $h(t,r) =
f_{1}(t) + r^{2}$ as Type $B$ solutions.  When $f_{1}(t) = Const.$, say,
$f_{1}$, we can introduce a new radial coordinate $\bar{r}$ via the relation
\bq
\lb{eq19}
\bar{r} = \frac{r}{f_{1} + r^{2}},
\eq
then the metric (\ref{eq2}) becomes
\bq 
\lb{eq20} 
ds^{2} = G(t, r) \left( dt^{2} -  \frac{{d\bar{r}}^{2} }{1 -
4f_{1}\bar{r}^{2}} - \bar{r}^{2}d\Omega^{2}_{N-2}\right),\;\;\;
\left(f_{1}(t) = Const.\right).
\eq 
If we further set $G(t,r) = G(t)$, the above metric becomes the 
Friedmann-Robertson-Walker (FRW) metric but in N-dimensional spacetimes
\cite{CB1990}. When $f_{1} = 0$, the solutions will reduce to the ones of
Type A.

In order to solve the Einstein
field equations (\ref{eq3}) - ({\ref{eq7}) or (\ref{eq8}) - ({\ref{eq12}),
we need to specify the equation of state for the fluid, which we shall
take as $p = k \rho$, where $k$ is an arbitrary constant.
Then, we find two classes of solutions, one has the CSS symmetry, while
the other does not. These solutions  are given as follows:

{\bf Type $A$ solutions}.  In this case, the solutions are given by
\bq 
\lb{eq21} 
h(t, r) = 1,\;\;\;\;  G(t,r)  =  \left(1 - Pt  \right)^{2 \xi}, 
\eq  
where $P$ is a constant, which characterizes the strength of the spacetime
curvature. In particular, when $P = 0$, the spacetime becomes Minkowski. The
constant $\xi$ is a function of $k$ and the spacetime dimension $N$, given by  
\bq 
\lb{eq22} 
\xi \equiv  
\frac{2}{(N-3)+(N-1)k}. 
\eq 
The corresponding energy density and velocity of the fluid are given,
respectively, by
\bqn
\label{eq23}
p &=& k\rho  =  {3k \xi^2 P^2}{\left( 1 - Pt  \right)^{-2(\xi+1)}  },\nb\\
u_0 & =& {\left(1 - Pt  \right)^{\xi}  },
\; \; \; \; \; \; u_1  = 0.
\eqn
When $k = 0, (N-1)^{-1}$, the above solutions reduce, respectively, to the one
for a dust and radiation fluid, studied in the context
of higher dimensional cosmology \cite{CB1990}. Except for these two
particular cases, the solutions, to our knowledge, are new. In the following,
we shall use them to model the gravitational collapse of a compact
ball, and   leave the study of their cosmological implications, together with
the one of Type B solutions to be given below, be considered in
\cite{RW1999}. 
  
{\bf Type $B$ solutions}. In this case, the condition $G(t,r) = G(t), \;
f_{1}(t) = Const.$ leads to the following solutions, 
\bq
\lb{eq24}
h(t, r) = \frac{1}{f_{1} + r^{2}},\;\;\; G(t)  = 
\left[A \cosh(\omega t)
+ B \sinh(\omega t) \right]^{2\xi},
\eq
where $ \omega \equiv  2 \sqrt{-f_1}/ \xi$, $A$ and $B$ are integration
constants, and $\xi$ is  defined by (\ref{eq21}).
 The energy density and
velocity of the fluid now are given by
\bqn
\label{eq25}
p &=& k \rho  =  
12kf_{1}(A^{2} - B^{2})\left[A \cosh(\omega t) 
+ B \sinh(\omega t) \right]^{-2(1+\xi)}, \nb\\
u_0 & = & \left[A \cosh(\omega t) 
+ B \sinh(\omega t) \right]^{\xi},\;\;\; u_{1} = 0.
\eqn
Clearly, this class of solutions also belongs to the FRW family, but with the 
curvature of the (N-2)-unit sphere different from zero. In fact, when $f_{1}
> 0$, the curvature is positive, and the spacetime is close, and when $f_{1}<
0$, the curvature is negative, and the spacetime is open. As far as we know,
these solutions are new.    

It should be noted that the above solutions are valid for any constant $k$.
However, in the rest of the paper we shall consider only the case where $ 0 \le
k \le 1$, so that the dominant energy condition is satisfied \cite{HE1973}.
When $N = 4$, the solutions reduce to the FRW solutions, which have been
studied in the context of gravitational collapse in \cite{WRS1997,RWS1999}.
Therefore, in the following we shall assume that $N \not= 4$.

\section {Gravitational Collapse of Perfect Fluid in N-dimensional Spacetimes}

To study the above solutions in the context of gravitational collapse, we
need first to define the local mass function. Recently, Chatterjee
and Bhui generalized the Cahill and Macvittie  mass function 
in four-dimensional spacetimes \cite{CM1973} to N-dimensional spacetimes
\cite{CB1993},  
\bq
\lb{eq26}
m_{CB}(t,r) = \frac{(N-3)r^{N-3}_{ph}}{2}R^{3}_{232},
\eq
where $r_{ph}$ is the physical radius defined by $r_{ph} \equiv rhG^{1/2}$
for the metric (\ref{eq2}). It can be shown that this definition is
consistent with, but not equal to, the following one, 
\bq
\label{eq27}
1- \frac{2 m(t,r)} { B_{N}r_{ph}^{N-3}} = -g^{\mu \nu}{r_{ph}}_{,\mu}
{r_{ph}}_{,\nu},
\eq
where 
\bq
\lb{eq28}
B_{N} = \frac{\kappa \Gamma\left(\frac{N-1}{2}\right)}{2(N-2)\pi^{(N-1)/2}},
\eq
with $\Gamma$ denoting the gamma function.
Clearly, when $N = 4$, it reduces to the one usually used in four-dimensional 
spacetimes \cite{PI1990}, and when the spacetime is static, it will give the
correct mass of black holes in N-dimensions \cite{MP1986}. Thus, in this
paper we shall use Eq.(\ref{eq27}) as the definition for the mass
function, from   which we can immediately localize the apparent
horizons, which are given by 
 \bq 
\lb{eq29}  g^{\mu \nu}{r_{ph}}_{,\mu}
{r_{ph}}_{,\nu} = 0. 
\eq
Then, the mass function on the apparent horizon is given by
\bq 
\lb{eq30} 
 M_{AH} = \frac{B_{N}}{2} r_{AH}^{N-3},
\eq 
which is usually taken as the  mass of black holes in gravitational collapse.  
With the above definition for the mass function, let us study the main
properties of the above two types of solutions separately.

\subsection{ Type A solutions }

The mass function definded by Eq.(\ref{eq27}) in this case takes the form
\bq
\label{eq31}
m(t,r) = \frac{B_{N}}{2}\frac{\xi^{2}P^{2}r^{N-1}}{(1 - Pt)^{2 - \xi(N-3)}},
\eq
while Eq.(\ref{eq29}) has the solution  
\bq
\label{eq32}
r_{AH} = {1 \over \xi} {| 1- Pt | \over | P |},
\eq
which represents the location of the apparent horizon of the solutions. When 
$\xi = 1 $,  the apparent horizon
represents a null surface in the $(t,r)$-plane, and when $ 0 \le \xi <
1$, the apparent horizon  is spacelike, while  
when $   \xi > 1$, it is  timelike. The spacetime is singular when
$t = 1/P$. This can be seen, for example, from the Kretschmann scalar, which
now is given by
\bq
\label{eq33}
{\cal{R}} \equiv
R^{\alpha\beta\gamma\delta} R_{\alpha\beta\gamma\delta}
  = 
 6\xi^{2} P^{4}\left\{N- 1+\xi^{2}
\left[ N- 2 + \sum_{A=1}^{N-3} (N-2-A) \right] \right\}
(1 - Pt)^{-4(1+\xi)}.
\eq
When $P > 0$, it can be shown that the singularity always hides behind the
aparent horizon, and when $P < 0$, the singularity is naked. In the latter
case, the solutions can be considered as representing cosmological models,
while in the former the solutions as representing the formation of black holes
due to the gravitational collapse of the perfect fluid. 
Substituting Eq.(\ref{eq32}) into Eq.(\ref{eq30}) we find that, as $t
\rightarrow + \infty$, the mass of the black hole becomes infinitely large. To
remend this shortage, in the next section we shall cut the spacetime along
a timelike hypersruce, say, $r = r_{0}$, and then join the part
with $r < r_{0}$ with an asymptotically flat Vaidya solution in N-dimensional
spacetimes. 

It is interesting to note that this class of solutions admits a homothetic
Killing vector,
\bq 
\label{eq34} 
\zeta_{0}  =  - \frac{1- Pt}{ (1 +  \zeta) P},  \; \; \; \; \; \; 
\zeta_1    =   \frac{r}{ 1 +  \zeta}, 
\eq 
which satisfies the conformal Killing equation, 
\bq
\lb{eq35}
\zeta_{\mu;\nu} + \zeta_{\nu;\mu} = 2 g_{\mu\nu}.
\eq
Introducing two new coordinates via the relations,
\bq
\lb{eq36}
\bar{t} = \frac{(1 - Pt)^{\xi + 1}}{(1 + \xi)P},\;\;\;
\bar{r} = r^{1 + \xi},
\eq
we find that the metric  can  be written  in an explicit self-similar form,
\bq  
\label{eq37} 
ds^2  =    {d\bar t}^2 -   
\left[ { {(\xi +1)}^{-1/\xi} P x} \right]^{2\xi \over \xi +1} d{\bar r}^2 
- \left[ {  {(\xi +1)} P} x \right]^{2\xi \over \xi +1}  
{\bar r}^2 d\Omega^{2}_{N-2}, 
\eq
where $x \equiv \bar{t}/\bar{r}$.

It is well-known that an irrotational 
``stiff" fluid ($k = 1$) in
four-demensional spacetimes is  energetically equal to a massless scalar
field \cite{TT1973}. It can be shown that this also the case for N-dimensional
spacetimes. In particular, for the above solutions with $k = 1$,   the
corresponding massless scalar field $\phi$ is given by 
\bq   
\lb{eq38}  
 \phi  =\pm \left[\frac{N-1}{\kappa (N-2)}\right]^{1/2}\ln\left(1-Pt\right)
+\phi_{0},  
\eq 
where $\phi_{0}$ is a constant.

\subsection{ Type B solutions} 

The solutions in this case are given by  (\ref{eq24}). When
$f_{1} > 0$, the spacetime is close, and  to have the metric real
 the constant $B$ has to be imaginary.  The spacetimes are singular when,
\bq
\lb{eq39}
t|_{f_{1} > 0} = \frac{1}{|\omega|}{\mbox{arctan}}\left(\frac{|B|}{A}\right) +
2n\pi, \eq
where $n$ is an integer. When $f_{1} < 0$, the spacetime is singular only
when
\bq
\lb{eq40}
t|_{f_{1} < 0} = \frac{1}{\omega}{\mbox{arctanh}}\left(\frac{B}{A}\right).
\eq
Therefore, in the following we shall consider only the case where $f_{1} < 0$.
In this case, to have the energy density of the fluid be non-negative, we
need to impose the condition $B^{2} \ge A^{2}$. Then, the metric coefficent
$G(t)$ can be written as 
\bq
\lb{eq41}
G = \left(B^{2} - A^{2}\right)^{\xi}
 \sinh^{2\xi}[\omega(t_{0} - \epsilon t)],
\eq
 where  $\epsilon = sign(B)$, and $t_{0}$ is defined as 
$$
\sinh(\omega t_{0}) = \frac{A}{(B^{2} - A^{2})^{1/2}}.
$$
Clearly, the conformal factor $(B^{2} - A^{2})^{\xi}$ does
not play any significant role, without loss of generality, in the following we
shall set it to be one. If we further introduce a new radial coordinate via the
relation,
\bq
\lb{eq42}
\bar{r} = - \int{h(t,r) dr} = \frac{1}{a}\ln\left|\frac{a + r}{a - r}\right|,
\eq
where $ a \equiv (-f_{1})^{1/2}$, the corresponding metric takes the form,
\bq
\lb{eq43}
ds^2 = 
{\sinh}^{2 \xi} \left[ {2 a\over \xi} \left( t_0 -\epsilon t \right) \right]
\left\{ dt^2 - d{ r}^2 - {{\sinh}^2(2  { r})\over 4 ^2}
d\Omega^{2}_{N-2} \right\}. 
\eq
Note that in writing the above expression, we had dropped the bars from 
$r$. Then, the mass function and Kretschmann scalar are given,
respectively, by
\bqn 
\lb{eq44}
m(r,t) &=&   {B_{N} \over 2^{N-2} } \sinh^{N-1}(2  {r} )
\sinh^{\xi(N-3) -2} {\left[ {2\xi^{-1}} {( t_0 - \epsilon t)} 
\right]}, \nb\\
{\cal{R}}  &=&   \frac{96}{\xi^2} \left\{ (N-1) +\xi^2
\left[ N-2 + \sum_{A=1}^{N-3} (N-2-A) \right] \right\} 
\times 
\sinh^{-4 (\xi +1)}  
{\left[ {2 \xi^{-1}} {\left( t_0 - \epsilon t\right)} 
\right]},
\eqn
while the apparent horizon is located at
\bq
\lb{eq45}
r = r_{AH} \equiv { \xi^{-1} } (t_0 - \epsilon t).
\eq
From the above equations, we can see that  the solutions
are singular  on the hypersurface $t =  \epsilon t_{0}$. When $
\epsilon = +1$, the   singularity is hidden behind the apparent horizon, and
the solutions  represent  the formation of black holes from the gravitational
collapse of the fluid. When $ \epsilon =  - 1$,  the singularity is naked, the
solutions can be considered as representing cosmological models or   white
holes.  As in the type A case, the mass   of such formed black holes also
diverges at the    limit $t \rightarrow + \infty$. Thus, to have finite masses
of black holes, we also need to make a ``surgery"  to the spacetimes. This
will be considered, together with the Type A case, in the next section. 

Before closing this section, we note that, when $k  = 1$, the corresponding
massless scalar field is given by
\bq
\lb{eq46}
\phi = \pm
\left[\frac{N-1}{\kappa(N-2)}\right]^{1/2}\ln\left\{\tanh[(N-2)(t_{0} -
\epsilon t)]\right\} + \phi_{0},\; (k = 1). 
\eq

\section{ Matching the Solutions with An Outgoing Radiation Fluid}

In order to have the black hole mass finite, one way is to cut
the  spacetimes  along a timelike hypersurface, say, $r = r_{0}(t)$, and then
join the internal part with an asymptotically flat spacetimes. From
Eqs.(\ref{eq23}) and (\ref{eq25}) we can see that the fluid is comoving in both
of the two cases. Thus, the timelike hypersurface now should be given by 
\bq
\lb{eq46a}
r = r_{0} = Const.
\eq
Then, from Eq.(\ref{b7}) we find that the extrinsic curvature is given by
\bq
\lb{eq47}
K_{\tau\tau}^- = 0,\;\;\;
K_{22}^- = \left\{ \prod_{k=2}^{i-1}\sin^2\left(\theta^k\right) \right\}^{-1}
K_{ii}^{-} = {J^\prime(r_{0}) J(r_{0}) }{F^{\xi}(t)},\; (i = 2,3,4, ..., N-1),
\eq
where a prime denotes the ordinary differentiation, and 
\bqn
\lb{eq48}
F(t) &=& \left\{
\begin{array}{ll}
1- Pt,& {\mbox{Type A}},\\
\sinh[2a\xi^{-1}(t_{0} - \epsilon t)],& {\mbox{Type B}},
\end{array} \right. \nb\\
J(r) &=& \left\{
\begin{array}{ll}
r,& {\mbox{Type A}},\\
\frac{1}{2}\sinh(2r),& {\mbox{Type B}}.
\end{array} \right. 
\eqn
There are various possiblities to choose the solutions outside the hypersurface
$r =r_{0}$. In this paper we shall choose the out-going Vaidya solutions in
N-dimensions \cite{CBB1990},
\bq
\lb{eq49}
ds^{2} =   \left[
 1- \frac{2m(v)}{B_{N}R^{N-1}}\right]dv^2 +2dvdR^2 -R^2d\Omega^{2}_{N-2}, 
\eq
which is a particular case of Eq.(\ref{b8}). The hypersurface $r = r_{0}$
in these coordinates is given by $R = R_{0}(v)$, or
\bq
\lb{eq49a}
R = R(\tau),\;\;\; v = v(\tau),
\eq
where $\tau$ is defined via the relation
\bq
\lb{eq49b}
d\tau = \left[1- \frac{2M(\tau)}{B_{N}R^{N-1}} + 2\frac{dR}{dv}\right]^{1/2}dv,
\eq
where $M(\tau) \equiv m(v(\tau))$. This equation can be also written as, 
\bq
\lb{eq51}
M (\tau) \; = \; \frac{B_{N}}{2}\frac{R^{N-3}}{\dot{v}^{2}}
\left(\dot{v}^{2} + 2\dot{v}\dot R - 1\right).
\eq
Then, from Eq.(\ref{b14}) we find that the
extrinsic curvature in these coordinates is given by 
\bqn 
\lb{eq50}
K_{\tau\tau}^+ &=& \frac{\ddot v}{\dot v} - 
 \frac{{\dot v} M(\tau)}{R^{N-2}},\nb\\
K_{22}^+ & = &
\left\{\prod_{k=2}^{i-1}\sin^2\left(\theta^k\right)\right\}^{-1}  K_{ii}^+
=  R\left\{ 
{\dot v}\left[ 1- \frac{2M(\tau)}{ (N-3)R^{N-3}}\right]  + \dot{R} \right\},
\; (i =2,3,4, ..., N-1).
\eqn 
Hence, the surface energy-momentum tensor, defined by
\cite{Israel1966}, 
\bq
\lb{eq52}
\tau_{AB} =  - \frac{1}{\kappa}
\left\{ \left[K_{AB}\right]  - g_{AB}  \left[K\right] \right\}, \; (A,B = 1,
2, ..., N-1),
 \eq
where $[K_{AB}] \equiv K^{+}_{AB} - K^{-}_{AB},\; [K] = g^{AB}[K_{AB}]$,
now can be written in the form,
\bq
\lb{eq53}
\tau_{AB} = \sigma w_A w_B  + 
\eta \left[\sum_{i=2}^{N-1} \theta_{(i)A}\theta_{(i)B}\right],
\eq
where 
\bqn
\lb{eq54}
\sigma  &=&  \frac{2}{\kappa}
\left\{\dot{R} - \frac{1}{\dot{v}} + J'(r_{0})\right\}, \nb\\ 
\eta  & = & \frac{1}{2\kappa R \dot{v}}
\left\{ (N-3)  \dot{v}^{2} +(N-4){\dot{R}} - 2\ddot{v}R - 2\dot{v}J'(r_{0}) + 
(5-N)  \right\}, \nb\\
w_{A} &=& \delta^{\tau}_{A}, \;\;\;
\theta_{(i)A} = \left\{\prod_{j=2}^{i-1}\sin^2\left(\theta^j\right)\right\}
\delta^{i}_{A}, \; (i = 2, 3, ..., N-1).
\eqn
To fix the motion of the shell, we need to specify the equation of state of
the shell, which will be taken as $\eta  =  \alpha \sigma$, where $\alpha$
is an arbitrary constant. Then, Eq.(\ref{eq54}) yields,
\bq
\lb{eq55}
(N-3){\dot{v}}^2 - 2 R \ddot{v} + [N-4(1+\alpha)]\dot{v}\dot{R}
-2(1+2\alpha)J'(r_{0})\dot{v} + (5+4\alpha -N) = 0.
\eq
Unfortinately, we are not able to solve the above equation to get $v(\tau)$,
except for the cases where 
\bq
\lb{eq56}
\alpha =  \frac{N-4}{4}.
\eq
In order to have the shell satisfy the dominant energy condition \cite{HE1973},
$\alpha$ is restricted to $0\le \alpha \le 1$. Then, from the above
expression we can see that this is the case only when $ 4 \le N \le 8$. Since
the case where $N = 4$ has been already considered in \cite{WRS1997,RWS1999},
in the following we shall exclude this case.
 
\subsection{Type A solutions}

 In this case, it can be shown that 
 Eqs.(\ref{eq55}) and (\ref{eq56}) have the integral,
\bq
\lb{eq57}
 \dot{v}(\tau)  =   \frac{1-Y}{1-(N-3)Y},
\eq
where 
\bq
\lb{eq58}
x \equiv \left[(\xi + 1)(\tau_{0} -
\tau)\right]^{\frac{1}{\xi + 1}},\;\;\;
R_{0} \equiv r_{0}P^{\frac{\xi}{\xi + 1}},\;\;
Y\equiv e^{[(N-4)(2v_0R_0-x)/(2R_0)]},
\eq
and $v_{0}$ and $\tau_{0}$ are integration constants. 
Substituting the above expressions into Eq.(\ref{eq51}), we find that 
\bq
\lb{eq59}
 M(x)  =   \frac{B_{N}R_{0}^{N-3}}{2}\left\{{\left[ 2(N-4)-
\left((N-3)^2-1\right)Y\right]x Y 
 - 2(1-Y) [1-(N-3)Y]\xi R_0 }\right\}{(1-Y)^{-2}} x^{\xi(N-3) -1}.
\eq
At the moment $\tau = \tau_{AH}\; $(or $x = x_{AH} = \xi R_{0}$), 
the shell collapses inside the apparent horizon. Consequently, the 
total mass of such formed black hole is given by
\bq
\lb{eq60}
M_{BH} \equiv M(x_{AH}) =   \frac{B_{N}}{2}
\frac{\left\{ 4(N-3)+ \left[2-(N-2)^2  \right] Y_{ah}
 \right\} Y_{ah}  -2 }{\left( Y_{ah}- 1\right)^2 }  
 \left[\xi^{\xi}r_{0}^{(1+\xi)}\right]^{N-3}P^{(N-3)\xi},
\eq
where 
$
  Y_{ah}\equiv e^{(N-4)(2v_0-\xi)/2}.
$
Clearly, this mass  is finite and can be positive
by  a proper choice of the  parameter $v_{0}$  for any given $\xi$.
The contributions of the fluid and the shell to this mass are, respectively,
given by,  
\bqn
\lb{eq62}
m^{f}_{BH} &\equiv& m^{f}_{AH}(\tau_{AH}) = 
\frac{B_{N}}{2} {\xi^{\xi(N-3)}r_{0}^{(N-3)(1+\xi)}}P^{\xi(N-3)},\nb\\
m^{shell}_{BH} & \equiv & 4\pi B_{N}R^{N-2}(\tau_{AH})\sigma(\tau_{AH}) 
 = \frac{ 8\pi B_{N}\left(\xi^{\xi} 
r^{\xi+1} \right)^{N-3}
[ 1-(N-3)Y_{ah}]}{\kappa(Y_{ah}-1)} P^{(N-3)\xi}.
\eqn
From the above expressions we can see that all the  masses are 
proportional to $P$, the parameter that characterizes the strength
of the initial data of the collapsing ball. Thus, when the initial data is very weak
($P \rightarrow 0$), the mass of the formed black hole is very small
($M_{BH} \rightarrow 0$). In principle, by properly tuning the parameter
$P$ we can make it as small as wanted. Recall that now the solutions 
have CSS. 

Note that although the mass of black holes takes
 a scaling form in terms of $P$, the exponent $\gamma$
is not uniquely defined. This is because 
in the present case the ``critical" solution ($P = 0$)  
separates black holes from white holes,
and the latter is not the result
of gravitational collapse. Thus,  the solutions considered here
do not really represent the critical collapse. As a result,  
we can replace $P$ by any function $P(\bar P)$, and for each 
of such replacements, we will have a different $\gamma$ \cite{Gundlach1996}.
However, such replacements do not change the fact that by properly
tuning the parameter we can make black holes with masses as small
as wanted. It should be also noted that  the definition of the total
mass of a  thin shell is not unique.  It is possible to use
equally  other definitions, such as, $m^{shell}_{BH} 
\equiv M_{BH} - m^{f}_{BH} $, but our final conclusions will not depend on them.

\subsection{Type B solutions} 

In this case, the integration of Eq.(\ref{eq55}) yields,
\bq
\lb{eq63}
\dot{v}  =  \frac{1}{2(N-3)} \left\{ n_0 -n_1\tanh\left(t_1 + 
\frac{n_1}{2\sinh(2r_0)}t\right) \right\},
\eq
where
\bq
\lb{eq64}
n_0 \equiv (N-2)\cosh(2r_{0}),\;\;\;   
n_1 \equiv \left[(N-2)^2 \cosh^2(2 r_0)-4(N-3)\right]^{1/2}, 
\eq
and $t_{1}$ is an integration constant. This solution reduces to 
the ones studied in \cite{WRS1997,RWS1999}  when  $N = 4$.
It can be shown that now the total mass of the black hole is given by
\bqn
\lb{eq65}
M_{BH} &= &  \frac{ B_{N}{\rm{sinh}}^{N-3}(2r_0)
     \left\{\sinh\left[{2}{\xi}^{-1}(t_0-t)\right] 
     \right\}^{\xi(N-3)-1}}{2^{N-2}n_4^2}
       \left\{ n_4^2 \sinh\left[{2}{\xi}^{-1}(t_0-t)\right] 
  +  {\phantom{\frac{A}{A}}}  \right.  \nb \\  
& & - 4 (N-3)n_4  \cosh \left[t_1 + n_3 t\right] 
\sinh(2r_0)  \cosh^2 \left[t_1 + n_3 t\right] \left[ {2}{\xi}^{-1}(t_0-t) \right] 
 \nb\\
 & & \left. -4 (N-3)^2 n_4^2  
 \sinh(2r_0)   \sinh \left[ {2}{\xi}^{-1}(t_0-t) \right]
  {\phantom{\frac{A}{A}}}\right\},
\eqn
where
\bq  
 \lb{eq66}
n_3   \equiv  \frac{n_1}{2\sinh(2r_0)},\;\;\;
n_4    \equiv   n_0\cosh\left[ t_1+n_3 t \right] - n_1 
\cosh \left[ t_1+n_3 t \right].
\eq
The contributions of the collapsing fluid and shell to the total mass of
black hole are given, respectively, by  
\bqn
\lb{eq67}
M^{f}_{BH} &\equiv& m^{f}_{AH}(\tau_{AH}) = 
\frac{B_{N}}{2^{N-2}} \sinh^{(N-3)(\xi +1)}(2r_{0}),\nb\\
M^{shell}_{BH} &\equiv& 4\pi B_{N}R^{N-2}(\tau_{AH})\sigma(\tau_{AH}) 
 =\frac{\pi B_{N}\sinh^{\xi(N-3) +1}(2r_0)}{\kappa n_{4h}2^{N-1}} \nb \\
& & \times \left\{  n_{4h} \cosh(2r_0)
         - 2(N-3)\cosh \left[ t_1+n_3(t_0- \epsilon \xi r_0) \right]
    \right\},
\eqn
where $n_{4h} =  n_0 \cosh[t_1+n_3(t_0-\epsilon\xi r_0)].$
From the above expressions we can see that for any given $r_{0},\;
M_{BH},\; M^{f}_{BH}$ and $ M^{shell}_{BH}$ are always finite and non-zero.
Thus, in the present case black holes start to form with a mass gap.

\section{ Concluding Remarks}

The N-dimensional spherically symmetric spacetimes have been studied, and
the Christoffel symbols, the Riemann, Ricci and Einstein tensors have been
given explicitly in the three usually used systems of coordinates, the
Schwarzschild-like coordinates, Eddington-Finkelstein-like coordinates, and
the double null coordinates. We wish that this would simplify the studies of
these spactimes. The general form of the metric for conformally flat spacetimes
has been found. As an application of it, all the Friedmann-Robertson-Walker-like 
solutions for a perfect fluid with an equation of state  $p = k \rho$  are
found. Then, these solutions have been used to
model the gravitational collapse of a compact ball, by first cutting them 
along a timelike   hypersurface   and then joinng them with  asymptotically
flat Vaidya solutions in N-dimensional spacetimes \cite{CBB1990}. It has been
shown that when the collapse has continuous self-similarity, the formation of
black holes always starts with zero mass, and when the collapse has no such a
symmetry, the formation of black holes always starts with a finite non-zero
mass. This is consistent with our previous results obtained in four-dimensional
spacetimes \cite{WRS1997,RWS1999}. Thus, they  
provide further evidences to support the speculation that {\em the formation
of black holes always starts  with zero-mass for the collapse with
self-similarities, CSS or DSS}.  

Finally, we note  that none of these two classes of solutions studied in this
paper represent critical collapse, as the ``subcritical" solutions ($P <
0$) do not represent gravitational collapse, but rather white holes. Thus,
{\em whether the formation of black holes starts with zero mass or not is
closely related to the symmetries of the collapse (CSS or DSS), rather than to
that whether the collapse is critical or not}.

\section*{ Appendix A: The Christoffel Symbols, the Riemann, Ricci and Einstein
Tensors in N-Dimensional spherically symmetric Spacetimes}

\renewcommand{\theequation}{A.\arabic{equation}}
\setcounter{equation}{0}

The general metric for N-dimensional spacetimes with spherical symmetry
takes the form,
\bq
\lb{a1}
ds^2 = g_{ab}(x^{0},x^{1})dx^adx^b -C^{2}(x^{0},x^{1})d\Omega^{2}_{N-2},\;(a, b
= 0, 1). 
\eq
 
In this paper, we shall follow the convertions of d'Inverno \cite{DI1992},
except for the following:  Greek indices run from $0$ to $N-1$, lower-case
latin, $a, b, c, d, ...$, from $0$ to $1$, lower-case latin, $i, j, k, l,
...$, from $2$ to $N-1$, and upper-case latin, $A, B, C, D, ...$, from $1$ to
$N-1$. 

It can be shown that the non-vanishing Christoffel symbols, defined by 
\bq
\lb{a3} 
{}^{N}\Gamma^{\mu}_{\mu\lambda} = \frac{1}{2}{}^{N}g^{\mu\sigma}\left(
{}^{N}g_{\sigma\lambda,\nu} + 
{}^{N}g_{\nu\sigma,\lambda} - {}^{N}g_{\nu\lambda,\sigma}\right),
\; (\mu,\nu, \lambda = 0, 1, 2, ..., N-1),
\eq
are given by       
\bqn
\lb{a4}
{}^N\Gamma^{a}_{bc} &=& \Gamma^a_{bc},\;\;\;
{}^{N}\Gamma^{2}_{a 2}= {}^{N}\Gamma^{i}_{a i}
= \frac{C_{,a}}{C}, \nb\\ 
{}^{N}\Gamma^{a}_{2 2} &=& \left[\prod_{k =
3}^{i-1}{sin^2\left(\theta^{k}\right)} \right]^{-1}{}^{N}\Gamma^a_{i i}  =   C
C^{,a}, \nb\\ {}^{N}\Gamma^{2}_{3 3}
&=&
\left[\prod_{k = 3}^{i-1}{\sin^2\left(\theta^{k}\right)}\right]^{-1}               
         {}^N\Gamma^{2}_{ii}
= -\sin(\theta^{2})\cos(\theta^{2}),\nb\\  
{}^N\Gamma^{i}_{jj} (j > i) &=&  - \cos(\theta^{i})\sin(\theta^{i}) 
 \prod_{k= i+1}^{j-1}{\sin^2\left(\theta^{k}\right)}, \nb
\\ 
 {}^N\Gamma^i_{2 i}  &=&{\mbox{cotan}}(\theta^{2}),\;\;\;      
{}^N\Gamma^j_{ i j} (j > i) = {\mbox{cotan}}(\theta^{i}),
\eqn 
where $(\;)_{,a} \equiv \partial (\;)/\partial x^{\alpha},\; i, j = 2,3, 4,
..., N-1,\; a, b, c = 0, 1$, and $\Gamma^a_{bc}$ denote the Christoffel symbols
calculated from the two metric $g_{ab}$.  Note that in Eq.(\ref{a4}) the
repeating indices, one is up and the other is down, do not represent sum.  
The same for the cases where the two same indices are all down. 
 
The Riemmann tensor, defined by,
\bq 
\lb{a5}
{}^{N}R^{\mu}_{\nu\lambda\sigma} =   {}^N\Gamma^{\mu}_{\nu\sigma,\lambda} 
- {}^N\Gamma^{\mu}_{\nu\lambda,\sigma} 
+ {}^N\Gamma^{\mu}_{\delta\lambda}{}^N\Gamma^{\delta}_{\nu\sigma} 
-  {}^N\Gamma^{\mu}_{\delta\sigma}{}^N\Gamma^{\delta}_{\nu\lambda},
\eq
has the following non-vanishing components,
\bqn
\lb{a6}
{ }^{N}R_{abcd} &=& R_{abcd},\;\;\;
{ }^{N}R_{2323}  = -\sin\left(\theta^{2}\right)^2 C^2 \left(1  
+ C^{,a}C_{,a}\right),\nb\\
{ }^{N}R_{aibi} &=& 
\left[\prod_{k=2}^{i-1}{\sin^2\left(\theta^{k}\right)}\right]  C C_{;ab}, \nb\\
{ }^{N}R_{2i2i} &=& -
\left[\prod_{k=2}^{i-1}{\sin^{2}\left(\theta^{k}\right)}\right]
 C^2 \left(1 + C^{,a}C_{,a}\right),\nb\\ 
{ }^{N}R_{ijij}(j > i)
&=& - \left[\prod_{k=2}^{j-1}{\sin^{2}\left(\theta^{k}\right)}\right]
\left[\prod_{l=2}^{i-1}{\sin^{2}\left(\theta^{l}\right)}\right]  C^2\left(1 +
C^{,a}C_{,a}\right), 
 \eqn
where $C_{;ab}$ denotes the
covariant derivative with respect to the two metric $g_{ab}$, and  $R_{abcd}$
denotes the Riemann tensor calculated from this two metric, which has
only one independent component, say, $R_{0101}$,
\bq
\lb{a6a}
R_{abcd} =R_{0101}
\left(\delta^{0}_{a}\delta^{1}_{b}\delta^{0}_{c}\delta^{1}_{d}
- \delta^{0}_{a}\delta^{1}_{b}\delta^{1}_{c}\delta^{0}_{d}
-\delta^{1}_{a}\delta^{0}_{b}\delta^{0}_{c}\delta^{1}_{d}
+ \delta^{1}_{a}\delta^{0}_{b}\delta^{1}_{c}\delta^{0}_{d}\right).
\eq
Defined the Ricci tensor as, 
\bq
\lb{a7}
{}^{N}R_{\mu\nu} = {}^{N}R^{\lambda}_{\mu\lambda\nu},
\eq
we find that it has the following non-vanishing components,
\bqn
\lb{a8}
{ }^{N}R_{ab} &=& R_{ab} - (N-2)\frac{C_{;ab}}{C},\;\;\;
{ }^{N}R_{22} =  C\Box C + (N-3)\left(1+C^{,a}C_{,a}\right),\nb\\
{ }^{N}R_{ii} &=& \left[\prod_{k
=2}^{i-1}{\sin^{2}\left(\theta^{k}\right)}\right] { }^{N}R_{22},
\; (i = 2,3,4, ..., N-1),
\eqn
where $\Box C \equiv g^{ab}C_{;ab}$, and
 the 2-dimensional Ricci
tensor $R_{ab}$ in terms of $R_{0101}$ is given by, 
\bqn
\lb{a9}
R_{ab} \equiv R^{c}_{acb}
= R_{0101}\left[g^{00}\delta^{1}_{a}\delta^{1}_{b} - 
g^{01}\left(\delta^{0}_{a}\delta^{1}_{b} + \delta^{1}_{a}\delta^{0}_{b}\right)
+ g^{11}\delta^{0}_{a}\delta^{0}_{b}\right]. 
\eqn
Then, the Ricci scalar reads
\bq
\lb{a10}
{}^{N}R  = R - \frac{N-2}{C^2}\left[(2C\Box C + (N-3)\left(1
+C^{,a}C_{,a}\right) \right],  
\eq
where $R \equiv g^{ab}R_{ab}$, while the Einstein tensor, defined as
${}^{N}G_{\mu\nu} = {}^{N}R_{\mu\nu} - \frac{1}{2} {}^{N}g_{\mu\nu}{}^{N}R$,
has the non-vanishing components, \bqn
\lb{a11}
{}^{N}G_{a b} &=&  \frac{N-2}{2 C^2}
\left\{g_{ab}\left[2C\Box C + (N-3)(1 + C^{,a}C_{,a})\right] - 2
CC_{;ab}\right\},\nb\\ 
{}^{N}G_{22} &=&  - \frac{1}{2}
\left\{ C^{2}R + (N-3)\left[2C\Box C + (N-4)(1 + 
C^{,a}C_{,a})\right]\right\},\nb\\
 {}^{N}G_{ii} &=&  \left[
\prod_{k=2}^{i-1}{\sin^{2}\left(\theta^i\right)}\right]  {}^{N}G_{22},
\; (i = 2,3,4, ..., N-1).
\eqn
It can be shown that when $N=4$ the above expressions are consistent with the
corresponding ones given in \cite{PI1990}.

\section*{ Appendix B: The Extrinsic Curvature of a Timelike Hypersurface
           in N-Dimensional Spherical Spacetimes }

\renewcommand{\theequation}{B.\arabic{equation}}
\setcounter{equation}{0}

In this appendix, we shall give the extrinsic curvature of a
timelike hypersurface in three different systems of coordinates, which
are used very often in the laterature. They are the Schwarzschild-like
coordinates, Eddington-Finkelstein-like coordinates, and the double null
coordinates. In the following let us consider them separately.

\subsection{The Schwarzschild-like coordinates}
 
In these coordinates, the metric can be cast in the form,
\bq
\lb{b1}
ds^2= A^2(r,t)dt^{2} - B^2(r,t)dr^2 - C^2(r,t)d\Omega^{2}_{N-2}.
\eq

Then, the non-vanishing two-dimensional Christofell
simbols are given by
\bqn
\lb{b2}
\Gamma^t_{tt} &=&  \frac{A_{,t}}{A }, \; \; \; \;     
 \Gamma^r_{rr} =  \frac{B_{,r}}{B },\;\;\;    
\Gamma^t_{tr} =  \frac{A_{,r}}{A },\nb\\
 \Gamma^r_{tr} &=&  \frac{B_{,t}}{B },\;\;\; 
 \Gamma^t_{rr} =  \frac{B }{A^2 }B_{,t},\;\;\;
 \Gamma^r_{tt} =  \frac{A }{B^2 }A_{,r}.
\eqn
We also have 
\bqn
\lb{b3}
C^{,a}C_{,a} &=&
\frac{C^2_{,t}}{A^2}  - \frac{C^2_{,r}}{B^2},\nb\\
C_{;ab} &=&\left(C_{,t,t} - \frac{A_{,t} C_{,t}}{A} 
- \frac{A A_{,r} C_{,r}}{B^2}\right)\delta^{0}_{a}
\delta^{0}_{b}\nb\\
& & + \left(C_{,r,t} - \frac{A_{,r} C_{,t}}{A} - \frac{ B_{,t} C_{,r}
}{B}\right) \left(\delta^{0}_{a} \delta^{1}_{b} + \delta^{1}_{a}
\delta^{0}_{b}\right)\nb\\ 
& & + \left(C_{,r,r} - \frac{B_{,r} C_{,r}}{B} - \frac{B B_{,t} C_{,t}
}{A^2}
\right)\delta^{1}_{a} \delta^{1}_{b},\nb\\
\Box C   & = & \frac{1}{A^3 B^3} 
\left\{AB \left[ B\left( BC_{,t}\right)_{,t} - A\left( AC_{,r}\right)_{,r} 
\right] + \left( A^3 B_{,r}C_{,r} -B^3 A_{,t}C_{,t} 
\right) \right\},\nb\\
R_{0101}&=& AB\left\{\left(\frac{B_{,t}}{A}\right)_{,t}
- \left(\frac{A_{,r}}{B}\right)_{,r}\right\}.
\eqn
Substituting these expressions into Eq.(\ref{a11}) we find that the resulting
expressions for the Einstein tensor are consistent with those given in
\cite{CB1990}, except for the one of $G^{0}_{0}$, given by Eq.(5) in
\cite{CB1990}, where the third term $n\omega'/2$ should be replaced by
$n\omega''/2$.
 
For a timelike hypersurface,  
\bq
\lb{b4}
r = r_{0}(t),
\eq
the normal vector is given by 
\bq 
\lb{b5} 
n_\alpha = \frac{ABC}{[A^2-{r'}_{0}(t)B^2]^{1/2} } 
\left(-{r'}_{0}(t)\delta^t_\alpha +\delta^r_\alpha  
\right), 
\eq 
where ${r'}_{0}(t) \equiv dr_{0}(t)/dt$. On the surface, the metric
(\ref{b1}) reduces to 
\bq
\lb{b6}
ds^{2}|_{r = r_{0}(t)} = g_{AB}d\xi^{A}d\xi^{B} = 
d\tau^{2} - C^2\left(t,r_{0}(t)\right)d\Omega^{2}_{N-2}, 
\;(A, B = 1,2,3, ..., N-1), 
\eq
where the intrinsic coordinates are chosen as $\{\xi^{A}\}
 = \{\tau, \theta^{2}, \theta^{3}, ..., \theta^{N-1}\}$, with $\tau$ being
given by 
\bq
\lb{b6a}
d\tau =  \left[A^2-{r'}_{0}(t)B^2\right]^{1/2}dt.
\eq
Then, the extrinsic curvature, define by, 
\bq 
\lb{b6b} 
K_{AB} = - n_{\alpha}\left[ 
\frac{\partial^{2}x^{\alpha}}{\partial \xi^{A}  
\partial \xi^{B}} 
+ \Gamma ^{\alpha}_{\beta\delta}\frac{\partial  
x^{\beta}} 
{\partial \xi^{A}}\frac{\partial x^{\delta}} 
{\partial \xi^{B}}\right],\;  
\eq 
has the following non-vanishing components,
\bqn
\lb{b7}
K_{\tau \tau} & =& \frac{AB}{\left[A^2-{r'}_{0}(t)^2 B^2\right]^{1/2} } 
\left\{ {r'}_{0}(t)\left[ A{\ddot t} +{\dot t}^2 \frac{{A}_{,t}}{A}
+2 \dot t \dot r_0 \frac{{A}_{,r}}{A} + \frac{{\dot r_0}^2 B B_{,t} }{A^2} 
\right] \right. \nb\\   & &
{\phantom{ \frac{AB}{\sqrt{A^2-{r'}_{0}(t)^2 B^2} } }} \left.
 - \left[ {\ddot r_0} +{\dot r_0}^2 \frac{{B}_{,r}}{B}
+\dot t \dot r_0 \frac{{B}_{,t}}{B} + \frac{ {\dot t}^2 A  A_{,r}  }{B} 
\right] \right\} \nb\\   
 & =&  \frac{A B}{\left[A^2-{{r'}_{0}(t)}^2 B^2\right]^{3/2} } 
\left\{ {{r'}_{0}(t)}^3 \frac{B B_{,t}}{A^2} +{r'}_{0}(t)^2 
\left[ 2 \frac{A_{,r}}{A} - \frac{B_{,r}}{B} \right] \right. \nb\\
 & &
\left.+ {r'}_{0}(t) \left[ \frac{A_{,t}}{A} -2 \frac{B_{,t}}{B} \right]
- {r''}_{0} - \frac{A A_{,r}}{B^2}   \right\}, \nb\\
K_{22} & =&\frac{ABC}{\sqrt{A^2-{r'}_{0}(t)^2 B^2} }
\left[ \frac{ {r'}_{0}(t) }{A^2} C_{,t} + \frac{C_{,r}}{B^2} \right]
\nb\\ 
& = & \frac{C \dot C}{\dot r_0 B\sqrt{1-{\dot r_0}^2 B^2 }}\left[ 
 B^2 {\dot r_0}^2 +1 \right], \nb\\
 K_{ii} & =& \left[ \prod_{k=2}^{i-1}\sin^2\left(\theta^k\right) \right] 
 K_{22}, \; (i = 2,3,4, ..., N-1).
\eqn

\subsection{ The Eddington-Finkelstein-like coordinates}

In this case, the metric  can be written in the form
\bq
\lb{b8}
ds^2= e^{\psi(v,r)}dv
\left[ f(v,r)e^{\psi(v,r)}dv +2\epsilon dr \right]
-C^2(v,r)d\Omega^{2}_{N-2},
\eq
where $\epsilon = \pm 1$. When $\epsilon = + 1$, the radial coordinate $r$
increases toward the future along a ray $v = Const.$, i.e., the light cone 
$v = Const.$ is expanding.  When $\epsilon = - 1$, the radial coordinate $r$
decreases toward the future along a ray $v = Const.$, and the light cone 
$v = Const.$ is contracting. 

The two-dimensional non-vanishing Christofell simbols in this case are given
by  
\bqn
\lb{b9}
\Gamma^v_{vv} &=&  \psi_{,v}  -\frac{\epsilon}{2}e^{-\psi}
\left(e^{2\psi} f \right)_{,r},\;\;\;
\Gamma^r_{rr} = \psi_{,r},\nb\\
\Gamma^r_{vr}  &=& \frac{\epsilon}{2}e^{-\psi} \left(
e^{2\psi} f \right)_{,r},\; \; \;
 \Gamma^r_{vv} =  \frac{1}{2}
 \left[f\left(e^{2\psi}f\right)_{,r} +
\epsilon e^{\psi}f_{,v}  \right].
\eqn
Combining Eq.(\ref{a4}) and Eq.(\ref{b9}) given above with Eq.(4) in
\cite{CB1993}, we find that some non-vanishing terms of the Christoffel
symbols are missing in \cite{CB1993}.

We also have
\bqn
\lb{b10}
C^{,a}C_{,a} & = & 2e^{-\psi} C_{,r}C_{,v} - f {C_{,r}}^2 ,
  \nb\\
C_{;ab} &=& \left\{C_{,v,v} 
-\left[\psi_{,v} 
 - \frac{\epsilon}{2}e^{-\psi}
\left(e^{2\psi} f \right)_{,r}\right] C_{,v}
 -  \frac{1}{2}
 \left[ f\left(e^{2\psi}f\right)_{,r} + \epsilon e^{\psi}f_{,v}
 \right] C_{,r}
\right\}\delta^{v}_{a}\delta^{v}_{b}\nb\\
& & + \left[C_{,r,v}  - \frac{\epsilon}{2}e^{-\psi} \left( e^{2\psi}
f \right)_{,r}C_{,r}\right]\left(\delta^{v}_{a}\delta^{r}_{b}
+ \delta^{r}_{a}\delta^{v}_{b}\right)
+ \left(C_{,r,r} -  {\psi_{,r}}   C_{,r}\right)
\delta^{r}_{a}\delta^{r}_{b},\nb\\
\Box C &=& 2\epsilon e^{-\psi} C_{,r,v} - e^{-\psi}
\left(e^{\psi}f C_{,r}\right)_{,r},\nb\\
R_{0101} &=& e^{2\psi}\left\{ \epsilon e^{-\psi} \psi_{,vr} - f\psi_{,rr}
- f\psi_{,r}^{2} - \frac{3}{2}\psi_{,r}f_{,r} - \frac{1}{2}f_{,rr}\right\}.
\eqn
Inserting it into Eqs.(\ref{a6}) and (\ref{a11}), we find that the resulting
expressions for the Riemman and Einstein tensors are consistent with the
corresponding ones given in \cite{CB1993}.

A timelike hypersurface in these coordinates can be written as  
\bq
\lb{b11}
r  =  r_{0}(v),
\eq
or
\bq
\lb{b11a}
r  =  r(\tau),\;\;\; v = v(\tau).
\eq
On this hypersurface,  the metric will reduce to the one given by
Eq.(\ref{b6}) with $\tau$ being defined by
\bq
\lb{b12}
d\tau = e^{\psi/2}\left(fe^{\psi} +
2\epsilon\frac{dr_{0}(v)}{dv}\right)^{1/2}dv,
\eq
while its normal vector is given by 
\bq 
\lb{b13}
n_\alpha = e^{\psi}
\left( -\dot r\delta^v_\alpha + \dot v\delta^r_\alpha 
\right),
\eq
where $\dot{r} \equiv dr/d\tau$, etc. Then, it can be shown that the
extrinsic curvature tensor has the following non-vanishing components,
\bqn
\lb{b14}
K_{\tau \tau} & = & 
 \epsilon \frac{\ddot v}{\dot v} + \epsilon \dot{v}\psi_{,v}
- \frac{\dot{v}}{2}e^{\psi}f_{,r} - \dot{v}e^{\psi}f\psi_{,r},\nb\\
K_{22} & =& 
 C \left[ \epsilon\left({\dot r} C_{,r} - {\dot v} C_{,v}\right)
  +e^{\psi} f {\dot v} C_{,r} \right], \nb\\
& =& 
 C \left[ \frac{e^{-\psi}}{{\dot v} } \left( 1 + e^{2\psi} f {\dot v}^2
\right)C_{,r} - \epsilon{\dot v} C_{,v}    \right], \nb\\
 K_{ii} & =& \left[ \prod_{j=2}^{i-1}\sin^2\left(\theta^j\right) \right] 
 K_{22}, \; (i = 2,3,4, ..., N-1).
\eqn

\subsection{The Double Null  Coordinates}

In these coordinates, the metric can be written as,
\bq
\lb{b15}
ds^2= 2e^{2\sigma(u,v)}dudv-C^2(u,v)d\Omega^{2}_{N-2}.
\eq
Then, the two-dimensional non-vanishing Christoffel symbols are given by
\bq
\lb{b16}
\Gamma^v_{vv}  =  2 \sigma_{,v},\; \; \;  
 \Gamma^u_{uu}  =  2 \sigma_{,u},
\eq
while
\bqn
\lb{b17}
C^{,a}C_{,a} & = &
2e^{-2\sigma} C_{,u}C_{,v}, \nb\\
C_{;ab} & =& \left(C_{,v,v} - 2\sigma_{,v} C_{,v}\right)
\delta^{v}_{a}\delta^{v}_{b} 
+ C_{,u,v}(\delta^{v}_{a}\delta^{u}_{b} + \delta^{u}_{a}\delta^{v}_{b})
+  \left( C_{,u,u} -  2\sigma_{,u}   C_{,u}\right)
\delta^{u}_{a}\delta^{u}_{b},\nb\\
\Box C &= & 2 e^{-2 \sigma} C_{,v,u},\nb\\
R_{0101} &=& 2e^{2\sigma}\sigma_{,uv}.
\eqn
Substituting these expressions into Eqs.(\ref{a4}) - (\ref{a11}), we find that
the resulting expressions are consistent with the corresponding ones given in
\cite{WL1986} for the case $N = 4$.

A timelike hypersurface in these coordinates is given by,
\bq
\lb{b18}
u = u_0(v),
\eq
or 
\bq
\lb{b18a}
u = u(\tau),\;\;\; v = v(\tau).
\eq
On this hypersurface, the metric (\ref{b15}) also reduces to the one given by
Eq.(\ref{b6}) but now with $\tau$ being defined as,
\bq
\lb{b19}
d\tau = \left[2e^{2\sigma} \frac{du_{0}(v)}{dv}\right]^{1/2}dv, 
\eq
and its normal vector is given by
\bq
\lb{b20}
n_\alpha  = \frac{1}{ 2 {\dot u }}
\left( -2 e^{2 \sigma}{\dot u}^2 \delta^v_\alpha +\delta^u_\alpha \right).
\eq
It can be shown that in these coordinates the extrinsic curvature tensor
of the hypersurface has the following non-vnaishing components,
\bqn 
\lb{b21}
K_{\tau \tau}   & = & - \left(  2\sigma_{,u} \dot{u}
+ \frac{\ddot u}{\dot
u}\right), \nb\\ 
K_{22} & = &\frac{C}{2{\dot u} }\left(2C_{,v}\dot{u}^{2} -
e^{-2\sigma}C_{,u}\right),\nb\\
K_{ii} & =& \left[ \prod_{k=2}^{i-1}{\sin^2\left(\theta^{k}\right)} \right] 
 K_{22}, \;(i = 2,3,4, ..., N-1).
\eqn

\acknowledgments

We would like to express our gratitude to S.K. Chatterjee for
valubale discussions. The financial assistance from CAPES (JFVR), CNPq (AW)
and FAPERJ (AW) is gratefully acknowledged.

\end{document}